# Empowering Medical Imaging with Artificial Intelligence: A Review of Machine Learning Approaches for the Detection, and Segmentation of COVID-19 Using Radiographic and Tomographic Images

Sayed Amir Mousavi Mobarakeh, Kamran Kazemi, Ardalan Aarabi, Habibollah Danyali

***Abstract* — Since 2019, the global dissemination of the Coronavirus and its novel strains has resulted in a surge of new infections. The use of X-ray and computed tomography (CT) imaging techniques is critical in diagnosing and managing COVID-19. Incorporating artificial intelligence (AI) into the field of medical imaging is a powerful combination that can provide valuable support to healthcare professionals.This paper focuses on the methodological approach of using machine learning (ML) to enhance medical imaging for COVID-19 diagnosis.For example, deep learning can accurately distinguish lesions from other parts of the lung without human intervention in a matter of minutes.Moreover, ML can enhance performance efficiency by assisting radiologists in making more precise clinical decisions, such as detecting and distinguishing Covid-19 from different respiratory infections and segmenting infections in CT and X-ray images, even when the lesions have varying sizes and shapes.This article critically assesses machine learning methodologies utilized for the segmentation, classification, and detection of Covid-19 within CT and X-ray images, which are commonly employed tools in clinical and hospital settings to represent the lung in various aspects and extensive detail.There is a widespread expectation that this technology will continue to hold a central position within the healthcare sector, driving further progress in the management of the pandemic.**

***Index Terms*— COVID-19, Machine learning, Classification, Segmentation, Data acquisition .**

## I. Introduction

Since December 2019, the global population has encountered a profound health crisis in the shape of the COVID-19 pandemic, triggered by the SARS-CoV-2 virus. Despite the administration of over 13 billion vaccine doses as of December 13, 2023, new variants of the virus and mutations continue to contribute to the rising number of infections worldwide. As of May 3, 2023, the confirmed cumulative number of Covid-19 cases has reached 772,386,069, with 6,987,222 reported deaths in over 200 regions and territories [1]. During the recent 28-day period from April 3 to April 30, 2023, there were a total of approximately 2.8 million novel cases and in excess of 17,000 fatalities reported globally [2]. These findings suggest that while the majority of people have been vaccinated, the emergence of new variants can still lead to an increase in infections and fatalities. The COVID-19 disease exhibits various symptoms, including but not limited to cough, fever, shortness of breath, anosmia, and ageusia [3]. The polymerase chain reaction (PCR) technique is commonly perceived as the benchmark nucleic acid test for identifying specific viruses, and the real-time reverse transcriptase-PCR (RT-PCR) test is well known for its high sensitivity, specificity, and rapid detection [4]. However, the RT-PCR test has constraints due to kit performance and sample collection, with reported inadequate sensitivity ranging from 30% to 60% [5], [6]. In addition to RT-PCR, medical imaging techniques provide vital information for the diagnosis. It should be noted that the symptoms have not necessarily correlate with the severity of the infection [3]. When compared to RT-PCR, chest computed tomography (CT) exhibits greater sensitivity in detection [5]. CT scans are favored over X-rays owing to their ability to provide a three-dimensional depiction of the lungs and their wider use in the identification of pulmonary infections [7], [8], [9]. During the early stage, COVID-19 infections in CT scans are categorized as ground-glass opacity (GGO), while in the subsequent stage, they are characterized by pulmonary consolidation [5], [10]. GGO is a hazy area in the lungs that appears as ground glass on imaging scans, and it indicates that there is an abnormality in the air spaces of the lungs. Consolidation, on the other hand, occurs as a result of the infiltration of fluid, blood cells, and other substances into the air spaces in the lungs, causing them to become solid or ”consolidated”. Both GGO and consolidation are important indicators of COVID-19 infection and are often used in diagnostic and monitoring procedures.

The process of manually identifying lesions in lung CT scans is a laborious and repetitive undertaking for radiologists. Additionally, the accuracy of annotations is heavily reliant on the radiologist’s level of experience, with individual bias further impacting their precision. However, the emergence of machine learning (ML) in medical imaging has contributed significantly to improving the effectiveness of COVID-19 treatment protocols [11]. There exist various subcategories within machine learning, encompassing deep reinforcement learning, supervised learning, unsupervised learning, and semi-supervised learning [12]. Deep reinforcement learning is a technique that merges reinforcement learning with deep neural networks to enhance decision-making in intricate environments. One method of machine learning is supervised learning, which entails teaching a model using labeled data, while another method is unsupervised learning, which involves identifying patterns and relationships in unlabeled data. Semi-supervised learning integrates specific features and methods from both supervised and unsupervised learning approaches to enhance model precision on labeled data.

This comprehensive methodological review elucidates the substantial impact of integrating medical imaging and machine learning in the endeavor to overcome COVID-19. The significance of AI in medical imaging is particularly pronounced since COVID-19 was declared a pandemic. This review article summarizes the most popular and comprehensive machine learning approaches in segmentation, classification, and combination of classification and segmentation. Since machine learning-based models are mainly data-driven, several datasets have been introduced to facilitate the development of these models. The final section of this article discusses the remaining challenges and problems that researchers may encounter in the future. It is essential to note that the review primarily focuses on the most relevant methods up to May 5, 2023.

## II. Machine Learning for COVID-19 Classification

In regions experiencing COVID-19 outbreaks, the prompt identification and treatment of individuals suspected to have contracted the virus is of paramount importance. To this end, medical imaging modalities are commonly employed due to their expedited acquisition. However, medical images, particularly chest CT scans, often comprise A large number of image slices, necessitating significant time investments from radiologists for comprehensive analysis. Additionally, COVID-19 exhibits analogous characteristics to various forms of pneumonia, impeding radiologists' ability to attain high diagnostic precision without extensive expertise. Hence, there exists a pressing demand for AI-augmented diagnosis employing medical images. Subsequently, we shift our focus to methodologies that differentiate coronavirus cases from non-coronavirus cases and other forms of pneumonia infection in chest X-rays and CT scans.

### A. X-ray Imaging in Screening for COVID-19

While X-ray imaging is commonly employed as the initial diagnostic tool for COVID-19 in patients under investigation, it is widely recognized to have lower sensitivity in comparison to three-dimensional chest CT imaging. However, recent research suggests that early or mild cases of the disease may exhibit normal features in X-ray scans. According to a study, at the time of admission, 69% of patients had abnormal chest radiographs, which increased to 80% during hospitalization. COVID-19 is characterized by airspace opacities, GGO, and consolidation on radiographs, with the most frequently observed distributions being bilateral, peripheral, and lower zone predominant (90%). Abnormalities in the parenchyma are more prevalent than pleural effusion, which is uncommon (3%) [13]. Table I and Table II summarize the most relevant advanced research studies in this particular area.

*1) Classification of COVID-19 From Non-COVID-19:* Cutting-edge X-ray imaging methodologies have been innovated to discriminate COVID-19 patients from non-COVID-19 individuals with enhanced precision and accuracy. The latter includes common phenomena or normal individuals. An X-ray is a cost-effective approach in medical diagnoses [14]. Deep convolutional neural networks (CNNs) have made significant strides in medical image classification thanks to the extensive image datasets at a large scale. CNNs can learn hierarchical from local image data. However, handling irregularities in annotated data, particularly in real COVID-19 cases from X-ray images, Persist as a significant challenge.

Abbas *et al.* [15] has adapted the Decompose, Transfer, and Compose (DeTraC) approach. DeTraC consists of three main parts: Class Decomposition, Transfer Learning, and Class Composition. In the Class Decomposition stage, an ImageNet [16] pre-trained model is being used to create a deep feature space from inputs. The methodology involves first generating a condensed feature space from the input images, which is subsequently utilized to decompose the original classes into a new set of decomposed classes. To this end, in the Transfer Learning phase, the end-stage of classification layer of an ImageNet is customized to align with the decomposed classes, and the parameters of the adopted pre-trained CNN model are fine-tuned. Following this, in the Class Composition stage, the predicted labels are estimated for the decomposed classes, and the final classification is refined based on error-correction criteria. To highlight the efficiency and significance of the class decomposition layer in facilitating knowledge transfer through transfer learning, several pre-trained CNN models were employed as part of the study such as AlexNet [17], VGG [18], ResNet [19], GoogleNet [20], and SqueezeNet [21] and VGG19 exhibited the topmost accuracy in DeTraC. The accuracy of VGG-19 With class decomposition is 93.1%.

Ullah *el al.* [14] suggested a remedy for the shortage of annotated data through the implementation of a multi-task semi-supervised learning (MTSSL) fraHabibollah Danyali,mework. This framework involves incorporating auxiliary tasks for which there is an adequate amount of publicly available data. To further boost the effectiveness of multi-task learning, the authors also incorporated an adversarial autoencoder (AAE) into the MTSSL framework, which is capable of learning powerful and discriminative features.

Ali Narin*el al.* [22] utilized five pre-trained models, namely ResNet50 [23], ResNet101, ResNet152, InceptionV3 [24], and Inception-ResNetV2 for the purpose of detecting patients infected with coronavirus pneumonia through chest X-ray radiographs. The study aimed to conduct three separate binary classifications encompassing four classes. The assessment of the performance outcomes indicated that the ResNet50 model, following pre-training, attained the greatest classification accuracy, achieving a score of 98.0%.

Wang *el al.* [25] presented a detection model that incorporates a projection-expansion-projection-extension (PEPX) design, enabling human-machine collaboration in the detection process. This approach enhances representational capacity while significantly reducing computational complexity. The authors, also produced new datasets named COVIDx.

Rahimzadeh *el al.* [28] introduced a novel concatenated neural network that utilizes the Xception [26] and ResNet50V2 [27] architectures to classify chest X-ray images into three categories: normal, pneumonia, and COVID-19. The model was evaluated on a large dataset, yielding an average accuracy of 99.50% and a sensitivity of 80.53% specifically for the COVID-19 class. Across five folds, an overall accuracy of 91.4% was achieved.

Hyperparameters are essential components of machine learning algorithms, as they play a indispensable role in controlling the behavior of training algorithms and significantly impact the performance of resulting models. There are two primary approaches to hyper-parameter optimization: manual and automatic searching. Manual searching involves a labor-intensive process of manually searching for optimal hyperparameters, requiring expertise in the relevant field. However, even with expertise, manual searching can be insufficient when dealing with large datasets and numerous model parameters that require tuning, thereby necessitating the use of automatic searching methods [29], [30]. As a result, automatic search alternatives, such as grid search and random search algorithms, have emerged in the literature. However, both methods still face challenges such as the curse of dimensionality and the inability to efficiently perform time-consuming operations [29], [31]. Bayesian optimization stands as an algorithm that can handle such optimization problems by relying on an approximation technique based on prior knowledge [15], [28]. Bayesian optimization utilizes the outputs of model evaluations as observations to facilitate online learning. Consequently, the algorithm necessitates a training process to obtain a function learned from the available data.

Ucar *el al.* [32] have introduced a novel architecture named Deep Bayes-SqueezeNet-based COVIDiagnosis-Net with Bayesian optimization additive. SqueezeNet, a convolutional network, demonstrates remarkable performance advantages over AlexNet while employing a significantly reduced number of parameters.

Luz *el al.* [33] have introduced a novel approach for classifying COVID-19, normal, and pneumonia cases, employing the EfficientNet family as their model. Their model utilized a pre-trained EfficientNet, initially trained on the ImageNet dataset, and fine-tuned specifically for COVID-19 screening. The classification problem was evaluated using two approaches: flat subclass and hierarchical classification. The former treated classes independently, disregarding their relationships, while the latter considered the taxonomy of classes.

*2) Classification of COVID-19 From Other Pneumonia:* In clinical practice, the differentiation between common pneumonia, particularly viral pneumonia, and COVID-19 is essential since these

TABLE I: RELATED STUDIES ON X-RAY IMAGES FOR CLASSIFICATION OF COVID-19 FROM NON-COVID

| Literature | Subjects | Method | Task | Result | Highlight |
|---|---|---|---|---|---|
| Abbas *et al.* [15] | **Covid-19:** 105 images (4248 × 3480 pixels) **SARS:** 11 images (4248 × 3480 pixels) **Normal:** 80 images (4020 × 4892 pixels) | Decompose, Transfer, and Compose **(DeTraC)** | COVID-19 Normal | **VGG19** **Acc:** 93.1% **Sen:** 100% | Dealing irregularities in the image dataset |
| Ulla *et al.* [14] | **CheXpert Dataset:** **Pleural effusion:** 75,696 cases **Lung opacity:** 92,699 cases **Pneu:** 4,576 cases **COVIDx dataset:** **Normal:** 8,851 cases **COVID-19:** 1,770 cases **Pneu:** 6,069 cases | Multi-Task Semi-Supervised Adversarial AutoEncoding **(MTSS-AAE)** | Covid-19 Normal | **Acc:** 96.95% | |
| Narin *et al.* [22] | **COVID-19:** 682 cases **Normal:** 2800 cases **Vir. Pneu:** 1493 cases **Bac. Pneu:** 2772 cases | ResNet50, ResNet101, ResNet152, InceptionV3, Inception-ResNetV2 | Covid-19 Normal | **Resnet50** **Acc:** 99.7% | |
| Wang *et al.* [25] | 3,975 CXR images | COVID-Net | COVID-19 Pneu Normal | **Acc:** 93.3% **Sen:** 91% **PPV:** 98.9% | Reducing computational cost |
| Rahimzadeh *et al.* [28] | **COVID-19:** 149 images **Pneu:** 234 images **Normal:** 250 images | concatenation of the Xception and ResNet50v2 | COVID-19 Pneu Normal | COVID-19 class **Overall Acc:** 91.4% | |
| Ucar *et al.* [32] | **Normal:** 1203 cases **Pneu:** 1591 cases **COVID-19:** 45 cases | SqueezeNet | COVID-19 Pneu Normal | **Acc:** **COVID-19:** 100% **Normal:** 98.04% **Pneu:** 96.73% **Overall :** 98.26% | |
| Luz *et al.* [33] | 13,569 CXR images | EfficientNet | Covid-19 Pneu Normal | **Flat:** **Acc:** 93.9% **Sen:** 96.8% **Hierarchical:** **Acc:** 93.5% **Sen:** 80.6% | low computational cost |

**PPV:** Positive Predictive Values; **Sen:** Sensitivity; **Acc:** Accuracy; **Pneu:** Pneumonia; **Bac. Pneu:** Bacterial pneumonia; **Vir. Pneu:** Viral pneumonia conditions have similar radiological presentations. This distinction would be valuable in facilitating the screening process.
Khan *et al.* [34] Suggested architecture called CoroNet. CoroNet is based on the Xception architecture, and trained on the ImageNet dataset. The proposed methodology replaces traditional convolutions with depthwise separable convolution layers accompanied by residual connections. The introduced model addresses the challenge of vanishing gradients by incorporating residual connections. These connections enable the direct flow of gradients through the network, bypassing non-linear activation functions. As a result, the model mitigates the issue of gradient vanishing. The model achieved accuracy rate of 89.6% in the 4-class classification scenario, specifically for COVID-19 cases.

Hussain *et al.* [35] has proposed an alternative architecture, named CoroDet.CoroDet exhibits strong capability in accurately classifying X-ray and CT images into 2, 3, and 4 categories, namely COVID-19, normal, non-COVID viral pneumonia, and non-COVID bacterial pneumonia.

CoroDet achieved accuracies of 99.1%, 94.2%, and 91.2% for 2, 3, and 4-class classification problems, respectively.

Ozturk *et al.* [36] proposed altered version of Darknet-19 [37] called DarkCovidNet. YOLO (You only look once) [37] a system for object detection in real-time is ulilized as a start point for this architecture. The design of DarkCovidNet was influenced by the well-

TABLE II: RELATED STUDIES ON X-RAY IMAGES FOR CLASSIFICATION OF COVID-19 FROM OTHER PNEUMONIA

| Literature | Subjects | Method | Task | Result | Highlight |
|---|---|---|---|---|---|
| Khan *et al.* [34] | **Bac. Pneu:** 330 cases<br>**Vir. Pneu:** 327 cases<br>**COVID-19:** 284 cases<br>**Normal:** 310 cases | CoroNet | **4-Class:**<br>Bac. Pneu.<br>Vir. Pneu.<br>COVID-19<br>Normal<br><br>**3-Class:**<br>COVID-19<br>Pneu<br>Normal<br><br>**Binary:**<br>COVID-19<br>Normal | **4-Class:**<br>**Specificity:** 96.4%<br>**Acc:** 89.6%<br><br>**3-Class:**<br>**Spec:** 97.5%<br>**Acc:** 95%<br><br>**Binary:**<br>**Spec:** 98.6%<br>**Acc:** 99% | based on Xception |
| Hussain *et al.* [35] | **4-Class:**<br>**COVID-19:** 500 images<br>**Normal:** 800 images<br>**Vir. Pneu:** 400 images<br>**Bac. Pneu:** 400 images<br><br>**3-Class:**<br>**COVID-19:** 500 images<br>**Normal:** 800 images<br>**Bac. Pneu:** 800 images<br><br>**2-Class:**<br>**COVID-19:** 500 images<br>**Normal:** 800 images | CoroDet | **4-Class:**<br>Bac. Pneu<br>Vir. Pneu<br>Normal<br>COVID-19<br><br>**3-Class:**<br>Pneu<br>Normal<br>COVID-19<br><br>**Binary:**<br>COVID-19<br>Normal | **Acc:**<br><br>**2-Class:** 99.1%<br>**3-Cclass:** 49.2%<br>**4-Class:** 91.2% | |
| Ozturk *et al.* [36] | **COVID-19:** 125 images<br>**No-Findings:** 500 images<br>**Pneu:** 500 | DarkCovidNet | **Binary:**<br>COVID-19<br>NO.Finding<br><br>**3-Class:**<br>Pneu<br>COVID-19<br>NO.Finding | **Acc:**<br><br>**Binary:** 98.08%<br>**3-Cclass:** 87.02% | |
| Afshar *et al.* [38] | 94,323 CXR images | COVID-CAPS | Bacterial<br>COVID-19<br>Normal | **Acc:** 95.7%<br>**Sen:** 90%<br>**Spec:** 95.8% | |
| Gupta *et al.* [39] | **Covid:** 361 images<br>**Normal:** 1341 images<br>**Pneumonia:** 1345 images | InstaCovNet-19 | **3-Class:**<br>Covid-19<br>Pneumonia<br>Normal<br><br>**Binary:**<br>Covid-19<br>Normal | **Accuracy:**<br>**3-Class:** 99.08%<br>**2-Class:** 99.53% | |

**Spec:** Specificity; **Sen:** Sensitivity; **Acc:** Accuracy; **Pneu:** Pneumonia; **Bac. Pneu:** Bacterial pneumonia; **Vir. Pneu:** Viral pneumonia established DarkNet architecture. Instead of creating a new model from the ground up, the author made adjustments by reducing the number of layers and filters utilized in the architecture. The proposed model was specifically tailored to support both binary and multi-class classification tasks. The model attained a classification accuracy of 98.08% for binary classification and 87.02% for multi-class scenarios.

Afshar *et al.* [38] developed COVID-CAPS, a pioneering system that utilizes capsule networks to diagnose by analyzing 3D X-ray samples. Notably, this framework is applicable for small datasets. Experimental results demonstrated that COVID-CAPS achieved an accuracy rate of 95.7%.

Gupta *et al.* [39] has introduced a model called InstaCovNet-19, which utilizes various pre-trained models, namely ResNet101, Xception, InceptionV3, MobileNet [40], and NASNet [41], to compensate for the relatively small amount of training data. The objective of the proposed model is to detect COVID-19 and pneumonia by identifying the pathological abnormalities in Chest X-ray images that are characteristic of these diseases in affected individuals. To achieve this, the input images undergo concurrent processing by five pre-trained models, which operate in parallel mode.

The proposed model demonstrated an accuracy of 99.08% for three-class classifications and an accuracy of 99.53% for two-class classifications.

### B. CT Imaging in Screening for COVID-19

The radiological manifestations of COVID-19 on chest CT images have been systematically classified into four distinct stages, as documented in the existing literature [42]. During the initial stage, which typically occurs within the first four days of initial symptom manifestation, GGO are commonly observed in the lower lobes of the lungs. These opacities can be present unilaterally or bilaterally. In the advanced stage (within a timeframe of 5 to 8 days), there may be diffuse GGO, a pattern known as crazy paving, and consolidation, affecting multiple lobes on both sides of the lungs. The peak stage (within a timeframe of 9 to 13 days) is characterized by a higher prevalence of dense consolidation. Finally, during the absorption stage (typically occurring after a duration of 14 days), crazy paving patterns and consolidation steadily resolve, leaving behind only ground-glass opacities. These distinct radiographic patterns play a critical role in the CT-based classification and assessment of the severity of nifection. Table III and Table IV enumerate the most pertinent cutting-edge research conducted in this particular area.

*1) Classification of COVID-19 From Non-COVID-19:* In align- ment with the morphological features detected on X-ray images, sev- eral cutting-edge methodologies have been introduced to differentiate COVID-19 from other conditions, including cases of non-pneumonic and pneumonic pneumonia. Rahimzadeh *et al.* [43] designed the system to detect COVID-19 from high-resolution computed tomog- raphy (HRCT) scans of the lungs. In the initial stages of their study, the authors presented an algorithm that aims to effectively filter relevant images from patients' CT scans, ensuring accurate depiction of the lung interior. Nonetheless, when testing the neural network trained to classify COVID-19 cases using a specific dataset with visible lung interiors, there is a possibility of the network failing to detect cases where the lung appears closed at the initiation or termination of each CT scan image sequence. This is because the network has not been trained on such cases, resulting in incorrect detections and poor performance. To address this issue, the authors proposed dividing the dataset into three classes: infection-visible, no- infection, and lung-closed. However, this solution incurs additional costs, such as spending time creating new labels and changing the network evaluation method, and increases processing time as the network needs to evaluate all the images in the patient's CT scan. Alternatively, the proposed model utilizes techniques to discard images where the lung interior is not visible, reducing the processing time as the network only needs to evaluate selected images. The proposed model employs ResNet50V2 and a revised version of feature pyramid network (FPN) [44]. The FPN has been utilized in RetinaNet [45] to improve object detection.

the model demonstrated 98.49% accuracy for single image classification.

Wang *et al.* [46] presents a novel approach to classify COVID-19 cases using CT scans, addressing the challenges posed by limited data availability from a single site and the necessity for accurate classification models that can generalize to new data from diverse sites. To achieve this, the authors employ COVID-Net, a state-of-the-art deep learning architecture originally developed for COVID-19 CXR images, which has demonstrated superior performance compared to popular classification networks pre-trained on ImageNet. However, since COVID-Net is specifically designed for Chest X-ray(CXR) images with relatively coarse lesions, it may not effectively leverage the clearer lesion patterns and richer information present in CT images for optimal learning. Hence, the proposed method strives to enhance the learning efficiency and classification accuracy of COVID-Net by leveraging its strengths and incorporating batch normalization (BN) layers. The incorporation of BN layers in the model helps address the issue of internal covariate shift, leading to an enhanced feature discrimination capability and improved the rate of convergence. [47]. To tackle the challenge of obtaining a powerful and effective solution from heterogeneous COVID-19 datasets, the proposed method employs two approaches to enhance the diagnosis performance of COVID-Net. Firstly, the learning rate is Fine-tuned in a cosine annealing manner, allowing for a smoother and more effective exploration of the solution space. This helps the model to better adapt to the varying characteristics of different datasets. Secondly, two approaches are introduced to redesign COVID-Net, aiming to further improve its diagnosis performance. These modifications are intended to enhance the model's ability to accurately classify COVID-19 cases across diverse datasets. The first approach, called Separate Batch Normalization at Data Heterogeneity, tackles the issue of statistical discrepancies in heterogeneous COVID-19 datasets. The authors employ the domain-specific batch normalization (DSBN) method [48], [49], [50], where an individual batch normalization layer is assigned for each site Individually. This approach addresses the variations in statistics across different sites, helping the model to better adapt to the data heterogeneity. The second approach, known as Contrastive Domain Invariance Enhancement, utilizes contrastive learning. This method aims to enhance the model's ability to generalize across different domains by learning representations that are invariant to domain-specific variations. Contrastive learning [51] helps to capture the underlying patterns and common features across diverse datasets, thereby improving the model's robustness and performance.

Pathak *et al.* [52] utilized a deep transfer learning technique to detect the presence of infection. This approach involved training a pre-existing model on a large dataset and then transferring the learned knowledge to a new model designed for the specific task of detection. The proposed method attained a training accuracy of 96.22% and a testing accuracy of 93.01%, as reported by the author. Bai *et al.* [53] has introduced a novel technique that employs the EfficientNet B4 [54] architecture to differentiate between COVID-19 and various forms of pneumonia in individual patients. The methodology involves feeding abnormal CT slices, which have been segmented to isolate the lungs, into the EfficientNet B4 deep neural network structure. The model achieved a test accuracy of 96%, and when tested independently, it exhibited an accuracy of 87%. Radiologists were able to improve their average test accuracy to 90% by leveraging the model probabilities, surpassing their previous average accuracy of 85%.

*2) Classification of COVID-19 From Other Pneumonia:* Chest CT scans can exhibit similar characteristics in both COVID-19 infection and various forms of pneumonia. The most frequent findings observed in chest CT scans for COVID-19 infection are GGO and consolidation. However, these findings are not exclusive to COVID-19 and can also be observed in various forms of pneumonia.

Wang *et al.* [55] introduced a novel 2D model for the purpose of distinguishing between COVID-19 and typical viral pneumonia.

To enhance the model's performance, the authors incorporated transfer learning by training it with an existing model, specifically the GoogleNet Inception v3 CNN. The model achieved an accuracy of 89.5% on the testing dataset.

TABLE III: RELATED STUDIES ON CT IMAGES FOR CLASSIFICATION OF COVID-19 FROM NON-COVID

| Literature | Subjects | Method | Task | Result | Highlight |
|---|---|---|---|---|---|
| Rahimzadeh *et al.* [43] | **Normal:**48, 260 Slices<br>**COVID-19:** 15, 589 Slices | the feature pyramid network<br>**Backbone:** ResNet50V2 | COVID-19<br>other | **Acc:** 98.49% | Trained on large dataset |
| Wang *et al.* [46] | **Dataset-1:**<br>**COVID-19:** 1252 CT Slices<br>**Non-COVID:** 1230 CT Slices<br>**Dataset-2:**<br>**COVID-19:** 349 CT Slices<br>**Non-COVID:** 171 CT Slices | collaborative learning framework<br>**Backbone:** redesigned original COVID-Net | COVID-19<br>other | **Dataset-1:**<br>**Acc:** 90.83%<br>**Prec:** 95.75%<br>**Sens:** 85.89%<br>**Dataset-2:**<br>**Acc:** 78.69%<br>**Prec:** 78.02%<br>**Sens:** 79.71% | tackle the cross-site domain shift |
| Pathak *et al.* [52] | **COVID-19:** 413 CT Slices<br>**Non-COVID:**439 CT Slices | Deep Transfer Learning (DTL) | COVID-19<br>other | **Acc.** 93.01% | |
| Bai *et al.* [53] | 132,583 CT slices | EfficientNet B4 | COVID-19<br>other | **Acc:** 96%<br>**Sen:** 95%<br>**Spec:** 96%<br>**independent data:**<br>**Acc:** 87%<br>**Sen:** 89%<br>**Spec:** 86% | −320 HU used as the threshold value |

**Spec:** Specificity; **Sen:** Sensitivity; **Acc:** Accuracy; **Pre:** Precision

TABLE IV: RELATED STUDIES ON CT IMAGES FOR CLASSIFICATION OF COVID-19 FROM OTHER PNEUMONIA

| Literature | Subjects | Method | Task | Result | Highlight |
|---|---|---|---|---|---|
| Wang *et al.* [55] | **Non-COVID-19:** 740 images<br>**COVID-19 :** 740 images<br>**Vir. Pneu.:**740 images | CNN | Covid-19<br>Viral pneumonia | **Internal data:**<br>**Acc:** 89.5%<br>**spec:** 0.88%<br>**sen:** 0.87%<br>**External data:**<br>**Acc:** 79.3%<br>**Spec:** 0.83%<br>**Sen:** 0.67% | |
| Zhou *et al.* [56] | **Lung tumor:** 2500 cases<br>**Covid-19:** 2500 cases<br>**Normal:** 2500 cases | CNN Ensemble TL Majority Voting (EDL-COVID) | Lung tumor<br>covid-19<br>normal | **Acc:** 99.054<br>**Sen:** 99.05<br>**Spec:** 99.6 | |
| Ibrahim *et al.* [57] | **IAVP:** 224 cases<br>**COVID-19:** 219 cases<br>**Normal:** 174 cases | VGG-19 | COVID-19<br>Pneumonia<br>Lung cancer<br>Normal | **Acc:** 98.05%<br>**Spe:** 99.5% | |
| Polsinelli *et al.* [58] | **COVID-19:** 360 images<br>**Other illnesses and normal:** 397 images | SqueezeNet | COVID-19<br>CAP<br>non-pneumonia | **ACC:** 0.853%<br>**SEN:** 0.876%<br>**SPE:** 0.820%<br>**F1-score:** 0.86% | |

**Spec:** Specificity; **Sen:** Sensitivity; **Acc:** Accuracy; **CAP:** Community-acquired pneumonia; **IAVP:** influenza-A viral pneumonia

Zhou *et al.* [56] put forward an ensemble model called EDL-COVID, which combines three popular CNNs: AlexNet, GoogleNet, and ResNet18. The study focused on the clinical significance of lung CT images in the detection of COVID-19 and presented six relevant approaches. Detailed explanations and frameworks for transfer learning, ensemble models, and the individual CNNs (AlexNet, GoogleNet, and ResNet) were provided. An ensemble model then employed itself in combining the outputs from the three classifiers through relative majority voting, resulting in the conclusive classification output.The experimental results demonstrated that the ensemble model outperformed the individual models in terms of accuracy and detection time. The ensemble model achieved an optimized detection time while maintaining a classification accuracy of over 99%.

Ibrahim *et al.* [57] conducted an investigation to appraise the performance of four architectures for the classification of digital chest X-ray and CT datasets with four classes. The architectures studied in this research were VGG19-CNN, ResNet152V2, ResNet152V2 + Gated Recurrent Unit (GRU), and ResNet152V2 + Bidirectional GRU (Bi-GRU). The assessment involved the use of publicly available datasets. The experimental results indicated that the VGG19 + CNN model demonstrated superior performance in comparison to the other mentioned models. Specifically, the VGG19 + CNN model achieved an accuracy of 98.05%.

Polsinelli *et al.* [58] introduced the SqueezeNet architecture for the discrimination of COVID-19, Community-acquired pneumonia (CAP), and healthy cases using CT images. The proposed model demonstrates robust classification performance and operates efficiently on a moderate-speed computer without the need for GPU acceleration. Notably, the average classification time of the model is 1.25 seconds on a high-end workstation, which is highly competitive compared to more complex CNN models that require pre-processing, taking an average of 13.41 seconds. It is worth emphasizing that the proposed model can be executed on a mid-range laptop without requiring GPU-powered acceleration, achieving a completion time of 7.81 seconds. This is a significant achievement, as traditional GPU-accelerated methods often rely on specialized hardware. The model achieved an accuracy of 85.03%.

## III. MACHINE LEARNING FOR COVID-19 SEGMENTATION

Segmentation is an indispensable and fundamental step in image processing and analysis, as it involves the critical task of dividing an image into distinct and meaningful regions or objects for further examination and analysis. It involves the delineation of ROIs such as the lungs, lobes, bronchopulmonary segments, and affected areas or lesions in images. While CT imaging provides detailed 3D images that are effective for detection, X-ray imaging is more widely accessible globally. However, segmenting ROIs in X-ray images presents challenges on account of the presence of overlapping structures, such as ribs, which can obscure the visibility of soft tissues and complicate contrast.

Currently, there is a lack of established segmentation methods specifically for X-ray images in COVID-19 cases. However, advanced lung segmentation techniques designed for pneumonia cases in X-ray images can be adapted and utilized for COVID-19 diagnosis and other related diseases. Accurately segmenting ROIs in both CT and X-ray images is essential for precise assessment and comprehension of COVID-19 infection patterns. This segmentation process yields valuable information for diagnosis, monitoring disease progression, and devising suitable treatment strategies.

This subsection provides an overview of the state-of-the-art segmentation methods. These methodologies fall into two primary groups: lung-region-oriented and lung-lesion-oriented methods. Lung-region-oriented methods primarily focus on isolating the lung regions, including the entire lung and individual lung lobes, from the background regions. This step is considered essential in COVID-19 applications as it allows for subsequent analysis and assessment of lung abnormalities. On the other hand, lung-lesion-oriented methods specifically aim to separate the lesions associated with COVID-19 from the lung regions. Detecting and segmenting these lesions pose a challenge due to their small size, diverse shapes, and varied textures. Therefore, accurately locating and delineating these regions is considered a difficult task in COVID-19 segmentation. Several state-of-the-art studies have made significant contributions in this field. Table V and Table VI provide a all-inclusive list of the most relevant studies in this area. These studies highlight various segmentation techniques, algorithms, and evaluation metrics employed in COVID-19 segmentation.

### A. LUNG-REGIN-BASED

Lung-region-oriented segmentation methods are employed in the context of COVID-19 applications to specifically isolate the lung regions, such as the entire lung and its lobes, from the surrounding background areas. The primary objective of these segmentation methods is to accurately identify and delineate the boundaries of the lungs, thereby separating them from non-lung regions present in the images. This initial step of isolating the lung regions is crucial for subsequent analysis and detection of COVID-19-related abnormalities. Lung-region-oriented segmentation is indeed a pivotal step in analysis, as it lays the foundation for accurate diagnosis and quantification of the disease. This segmentation technique plays a vital role in identifying and isolating the lung regions in medical images, enabling subsequent analysis and detection of COVID-19-related abnormalities.

By accurately segmenting the lung regions, these methods pave the way for the detection of abnormalities and provide insights into the severity and progression of the disease. Lung-region-oriented segmentation has found widespread application in COVID-19 analysis, aiding in the identification and characterization of lung abnormalities caused by the virus. This technique contributes to the development of automated diagnostic tools and supports clinicians in making informed decisions regarding patient management and treatment strategies.

Punn *et al.* [59] Implemented the hierarchical segmentation network (CHS-Net) which was developed by drawing inspiration from various cutting-edge architectures, including U-Net [60], Google's Inception model, residual network, and attention strategy [61]. The approach employs two RAIU-Net models, which are connected in a series for hierarchical segmentation. The first model generates contour maps of the lungs, which are then used by the second model to identify infected regions. The proposed CHS-Net utilizes a multi-stage approach for feature extraction and segmentation. The method is partitioned into four phases, with each phase dedicated to extracting feature maps at distinct spatial dimensions. This multi-scale approach allows the model to capture both local and global information, enhancing its segmentation performance. The CHS-Net has been evaluated and achieved an accuracy of 95% in segmenting both the lung mask and the infection region.

Abdel-Basset *et al.* [62] proposed a few-shot U-Net network for segmenting similar infections from CT images. The model was trained with a Insufficient data and then refined by incorporating feedback from medical professionals on the segmented outputs.

Unlike traditional data-hungry image segmentation methods, this approach aims to achieve precise segmentation using a smaller labeled dataset. Typically, the dataset comprises only a few examples

TABLE V: RELATED STUDIES ON CT IMAGES FOR SEGMENTATION - LUNG-REGION-BASED

| Literature | Subjects | Method | Task | Result | Highlight |
|---|---|---|---|---|---|
| Punn *et al.* [59] | **Total:** 3560 CT images<br>**Covid-19:** 2200 CT images | CHS-Net | Lung lesion | **Lung segmentation:**<br>**Dice:** 96.3%<br>**Jaccard:** 94.7%<br>**Infection segmentation:**<br>**Dice:** 81.6%<br>**Jaccard:** 79.1% | |
| Asnawi *et al.* [63] | 20 CT scans of COVID-19 | 3D UNet | Lung lesion | **Lung Segmentation:**<br>**Dice:** 97.05%<br>**Infection Segmentation:**<br>**Dice:** 88.61% | CLAHE In preprocessing |
| Aswathy *et al.* [64] | 929 axial CT images | Cascaded 3D UNet | Lung lesion | **Lung segmentation:**<br>**Dice:** 92.46%<br>**Infection segmentation:**<br>**Dice:** 82.0% | |

TABLE VI: RELATED STUDIES WITH CT IMAGES FOR SEGMENTATION - LUNG-LESION-BASED

| Literature | Subjects | Method | Task | Result | Highlight |
|---|---|---|---|---|---|
| Allioui *et al.* [65] | 929 axial CT images | Deep Q-Network | Lung lobes Lesion | **Dice :** 80.81%<br>**Sen:** 79.97%<br>**Spec:** 99.48%<br>**F1 score:** 83.01% | |
| Fan *et al.* [7] | 100 axial Slices | Inf-Net | GGO consolidation | **Dice:** 68.2%<br>**Sen:** 69.2%<br>**Spec:** 94.3%<br>**Semi inf-net:**<br>**Dice:** 73.9%<br>**Sen:** 72.5%<br>**Spec:** 96.0% | utilizing a semi-supervised learning |
| Wang *et al.* [66] | **Training:** 52, 489 Slices<br>**Validation:** 6, 556 Slices<br>**Test:** 17, 205 Slices | COPLE-Net | Lesion | **Dice:** 80.72% | noise-robust Dice |
| Zhou *et al.* [67] | 201 CT scans (768 slices) | 2D U-Net | Lesion | **Dice:** 78.3% − 90.3% | |
| Zheng *et al.* [68] | **Total:** 4, 780 Slices<br>**Covid:**2, 506 Slices<br>**non-covid:** 2, 274 Slices | MSD-Net | GGO interstitial infiltrates consolidation | **Dice:**<br>**Interstitial infiltrates:** 73.84%<br>**Consolidation:** 87.69%<br>**GGO:** 74.22% | |

**Spec:** Specificity; **Sen:** Sensitivity; **Acc:** Accuracy; **GGO:** Ground-Glass Opacities

per class or category.
Asnawi *et al.* [63] employed the 3D UNet along with three other modified Frameworks derived from the 3D UNet: 3D ResUNet, 3D VGGUNet, and 3D DenseUNet. The experimental findings indicate that, despite having a significantly larger maximum iteration, the 3D UNet outperforms the three modified architectures, even when the latter were trained using transformer learning techniques. Aswathy *et al.* [64] have designed a two-tiered approach using a cascaded 3D UNet model to isolate infected areas in lung CT scans. The first stage involves applying a 3D UNet model to extract the lung parenchyma from the input CT volume. Preprocessing and augmentation techniques are employed to refine the data before feeding it into the model. The resulting lung parenchyma volumes are then post-processed to enhance the quality of the segmentation. In the second stage, another 3D UNet model is implemented to extract the infected volumes from the processed lung parenchyma. To diminish the presence of background pixels, only cropped parenchyma volumes are forwarded to the second 3D UNet model after number of processing. Given the 3D nature of the input data, significant GPU memory is required to perform the segmentation. To address this challenge, the approach utilizes a 3D patch-based strategy where the input volume is randomly divided into smaller patches, typically 16 patches, to optimize GPU memory usage during the segmentation process.

### B. LUNG-LESION-BASED

Lung-lesion-oriented segmentation refers to the process of identifying and segmenting abnormal or diseased regions within the lung parenchyma from medical images, typically CT scans. The goal of lung-lesion-oriented segmentation is to accurately delineate the boundaries of these lesions from the surrounding healthy tissue, which can be demanding task due to variations in size, shape, and appearance.

Allioui *et al.* [65] Put forward a novel to improve the segmentation by utilizing a multi-agent deep reinforcement learning approach. This method is an enhanced version of the DQN architecture and comprises agents that consist of two primary components: a CNN and a decision-making mechanism. The CNN extracts features from the CT images and Computes a probability distribution map that illustrates the probability distribution of each pixel belonging to the COVID-19-infected region. The decision-making mechanism determines the appropriate actions that the agents should take to optimize the segmentation performance. To promote effective collaboration among the agents, a reward system is introduced that rewards them based on their collaborative efforts towards achieving the common objective of producing an accurate segmentation map of the COVID-19-infected region in the CT images. The method was assessed using a publicly accessible dataset. The authors compared their results with those of other advanced segmentation techniques, such as U-Net++, COVNet, DeCoVNet, AlexNet, and ResNet.

Fan *et al.* [7] Presented a groundbreaking approach, Inf-Net, for segmenting COVID-19 lung infections. The method incorporates implicit reverse attention and explicit edge-attention mechanisms to boost the accuracy of identifying infected regions and enhance boundary representation. Additionally, the authors introduced Semi-Inf-Net, a semi-supervised approach to mitigate the deficiency of annotated data. The proposed approach has potential applications in COVID-19 diagnosis assessment, quantifying infected regions, assessing disease escalation, and analyzing screening data in bulk. Inf-Net demonstrates a remarkable capacity to detect objects characterized by a diminished intensity contrast between infected and normal tissues. Experimental results indicate that Inf-Net achieved a dice score of 0.739, while Semi-Inf-Net achieved a dice score of 0.682.
Segmenting large-scale 3D medical images poses a difficulty due to the presence of noisy labels, which can result from various factors such as poor contrast, and unclear boundaries. These challenges pose difficulties in accurately identifying and delineating the regions of interest in the images. In an effort to deal with this matter, Wang *et al.* [66] addressed the issue of learning from noisy labels They introduced a noise-robust Dice loss function that able to handle noisy labels and the Non-uniform distribution of foreground and background pixels. This loss function is compatible with different convolutional neural network architectures. Additionally, the authors developed a segmentation network called COPLE-Net. COPLE-Net incorporates several lightweight Components, inspired by U-Net variants, within an encoder-decoder structure.

Zhou *et al.* [67] The authors presented a novel method for segmenting and quantifying infection regions from various resources, addressing challenges such as limited data and model complexity. The authors introduced two innovative approaches to overcome these issues. Firstly, they developed a CT scan simulator specifically designed for COVID-19 by utilizing real patients' dynamic data at separate time instances. This simulator enabled the generation of synthetic CT scans, effectively augmenting the available data and mitigating data scarcity. Secondly, they proposed a unique deep learning algorithm that breaks down the 3D segmentation problem into three distinct 2D problems, significantly minimizing the overall model complexity. The segmentation task was divided into three planes: x-y, y-z, and x-z, and independent 2D U-nets were employed for each plane. This decomposition strategy not only improved segmentation accuracy but also enhanced computational efficiency.The results of the experiments demonstrated varying dice coefficients, a commonly used metric for evaluating segmentation accuracy, across different dataset scenarios. The range of reported dice coefficients, from 0.783% to 0.903%, signifies the effectiveness of the proposed method in accurately segmenting infection regions on CT scans. Zheng *et al.* [68] The authors introduced a novel approach called the Multi-Scale Discriminative Network (MSD-Net) for the segmentation of COVID-19 lung infection areas on CT scans. The lung infections were categorized into three types: ground-glass opacities, interstitial infiltrates, and consolidation. The MSD-Net incorporated three innovative blocks: the pyramid convolution block (PCB), channel attention block (CAB), and residual refinement block (RRB). The PCB utilized different kernel sizes and numbers to capture information at multiple scales, enabling the network to effectively segment infection areas of various sizes. The CAB combined features from different stages and emphasized the relevant information for accurate segmentation. The RRB further refined the extracted features to improve the segmentation results. In the experiments, the proposed MSD-Net demonstrated promising performance. The evaluation metric used, the Dice Similarity Coefficient, achieved values of 0.7422% for ground-glass opacities, 0.7384% for interstitial infiltrates, and 0.8769 for consolidation. These results indicate the effectiveness of the MSD-Net in accurately segmenting COVID-19 lung infection areas on CT scans.

## IV. MACHINE LEARNING FOR CLASSIFICATION AND SEGMENTATION

Numerous cutting-edge techniques have been put forth that integrate segmentation and classification. The primary rationale for employing this methodology is that, by utilizing the segmented lesion, the classification model can achieve greater diagnostic accuracy. Such methods have shown considerable efficacy in cases where classification models aim to differentiate between different types

TABLE VII: RELATED STUDIES WITH CT IMAGES FOR CLASSIFICATION AND SEGMENTATION LUNG-Lesion-BASED

| Literature | Subjects | Method | Task | Result | Modality |
|---|---|---|---|---|---|
| Bhattacharyya *et al.* [69] | **segmentation:** 247 CXR images **Classification:** **Normal:** 341 CXR images **Covid-19:** 342 CXR images **Pneumonia :** 347 CXR images | GAN CNN K-mean VGG DenseNet BRISK TLRF SVM XG Boost | **Segmentation:** lung **Classification:** Covid-19/ pneumonia / Normal | **Acc:** 96.60% **Spec:** 97.4% **Sen:** 95% | X-ray |
| Xu *et al.* [72] | **Normal :** 1840 **Covid-19:** 433 **tuber-culosis:** 394 **Bac. Pneu: :** 2780 **Vir. Pneu:** 1345 | GAN VGG ResUNet ResNet Inception V3 | **Segmentation:** lung **Classification:** Normal / Covid-19 / tuber-culosis / Bac. Pneu / Vir. Pneu | **Acc:** 96.03% − 97.06% **IOU:** 89.07% − 93.49% | X-ray |
| Degerli *et al.* [73] | 128,909 X-ray images | U-Net U-Net++ DLA DenseNet-121 CheXNet Inception-v3 ResNet-50 | **Segmentation:** Lung **Classification:** Covid-19/normal | **Segmentation:** **Sen :** 81.72% **Spec:** 99.93% **Acc :** 99.85% **Classification:** **Sen :** 94.96% **Spec:** 99.88% **Acc:** 99.73% | X-ray |
| Malhotra *et al.* [74] | **Segmentation:** **Lung mask:** 10567 **Disease mask:** 1707 **Classification:** **Covid-19:** 3963 **Normal:** 10250 **Other:** 20648 | COMiT-Net | **Segmentation:** Lung / disease localization **Classification:** Healthy / non-healthy / covid-19 / other disease | **Lung:** 0.85 **All Tasks:** **Sen:** 96.80 | X-ray |
| Ouyang *et al.* [75] | 4,982 CT images | 3D ResNet34 + an online attention module | **Segmentation:** Lung **Classification:** Covid-19/other | **Acc.**87.5% **Sens.**86.9% **Spec.** 90.1% | CT |
| Chen *et al.* [86] | 46,096 images | U-net ++ Backbone: Resnet-50 | **Segmentation:** lesion **Classification:** Covid-19/other | **Per Patient:** **Acc:** 95.24% | CT |
| Amyar *et al.* [87] | Covid-19: 449 Lung cancer: 98 Normal: 425 Other pathology: 397 | multi-task learning | **Segmentation:** Lesion **Classification:** Covid-19/normal/other | **Segmentation:** **Dice:** 88.0% **Classification:** **Acc:** 94.67% | CT |
| Gao *et al.* [88] | 1918 CT scans | DCN FCN U-Net | **Segmentation:** Lung/lesion **Classification:** Covid-19/normal | **Classification:** **ACC:** 96.74% **Lung Segmentation:** **DSC:** 99.11% **Lesion Segmentation:** **DCS:** 83.51% | CT |

**Sen:** Sensitivity; **Acc:** Accuracy; **Pneu:** Pneumonia; **Bac. Pneu.:** Bacterial pneumonia; **Vir. Pneu.:** Viral pneumonia

of pneumonia, rather than solely engaging in binary classification. Despite the absence of a well-established approach for segmenting covid-19 lesions in X-ray images, several techniques have been developed to segment the entire lung for the purposes of COVID-19 diagnosis. Table VII outlines the most pertinent cutting-edge research in area.

## A. X-ray Imaging in Screening for COVID-19

Bhattacharyya et al. [69] Presented a system that combines segmentation and classification techniques to classify COVID-19 X-ray images into normal, COVID, and pneumonia categories. The first step involved lung segmentation using a conditional GAN model, which helped identify ROI for further analysis. After successful segmentation, keypoint extraction techniques such as SIFT, BRISK, and K-means were employed to extract keypoints from the segmented lung regions. Next, deep learning models were applied to extract relevant characteristics from the extracted keypoints. These features captured important information that could distinguish between different categories. Finally, the extracted feature sets were used for the final classification of the X-ray images into normal, COVID, and pneumonia categories. By combining segmentation, keypoint extraction, feature extraction, and classification techniques, the proposed system aimed to accurately classify COVID-19 X-ray images based on the detected lung regions and extracted features. Goodfellow et al. [70] introduced Generative Adversarial Networks (GANs), a widely used model in image processing, for the purpose of translating an input image into its corresponding output image. GANs are made up of a generator and discriminator. The generator produces realistic images, while the discriminator categorizes them as fake or real. Through adversarial training, GANs generate high-quality and lifelike output images.For chest X-ray image segmentation, the Pix2Pix C-GAN model proposed by Isola et al. [71] was specifically used, where G stands as the generator and D stands as the discriminator. The former is the conventional encoder-decoder network and the latter is a patch-GAN encoder network.The proposed model outperformed five recent DL-based classifications with 96.6% accuracy.

Xu *et al.* [72] Formulated a network called Mask-Attention-based Neural Network (MANet) for COVID-19 X-ray classification. The study utilized a combination of three distinct datasets to create an input dataset with five class labels. The first step involved segmenting the lung regions from the X-ray images using a Residual U-Net (ResUNet) architecture.To perform the classification task, four different deep neural network models were evaluated: ResNet34, ResNet50, VGG16, and InceptionV3. Both versions of the models underwent testing, one with mask attention layers and one without. The mask attention layers aim to emphasize important regions in the X-ray images for better classification performance. The performance analysis revealed that the models with mask attention layers outperformed the models without them, with an approximate 2% improvement in performance. Among the tested models, the ResNet50 model with mask attention achieved the topmost classification accuracy of 97.06%.In terms of segmentation, the average test Intersection over Union (IoU) values for the normal, tuberculosis, and COVID-19 classes were reported as 93.49%, 93.14%, and 89.07%, respectively. These values indicate the accuracy of the segmentation results, with higher IoU indicating better overlap between the predicted and ground truth masks. Degerli *et al.* [73] introduced a detection and segmentation model and presented the ”Qata-COV19” dataset. The proposed model includes two-stage architecture, where manual segmentation applied to a randomly selected subset of the dataset, and models trained on this subset to generate segmentations. The best segmentations selected by medical doctors and used in stage 2 for training and cross-validation. Three different models, namely U-Net, U-Net++, and Deep layer aggregation (DLA), were employed with four pre-trained encoders (DenseNet-121, CheXNet, Inception-v3, and ResNet-50) in two settings: frozen or non-frozen encoder. All 24 combinations of these networks evaluated with two different dataset divisions to determine the best model. The U-Net and U-Net++ models with DenseNet-121 achieved the highest performance with over 99% accuracy for both taskes. Malhotra *et al.* [74] has presented an innovative architecture called COMiT-Net, which incorporates four simultaneous tasks. This model is inspired by the encoder-decoder architecture and includes two decoders: one for lung and disease segmentation. The lung segmentation decoder aims to generate well-defined boundaries encompassing the lung area. When evaluated using IoU scores, COMiT-Net demonstrates performance of 85%, while U-Net, Mask-RCNN, and SegNet achieve 82%, 85%, and 83%, respectively, for lung segmentation. Conversely, disease segmentation focuses on identifying abnormalities within the lung, such as unhealthy regions, COVID-19, and other diseases, rather than precisely delineating the boundaries of the disease.

## B. CT Imaging in Screening for COVID-19

Ouyang *et al.* [75] has put forward a approach for diagnosing COVID-19 from CAP. The approach involves an online attention module that focuses on infection regions in the lungs to aid in the diagnosis. The online attention module is an integral component of the method, designed to enhance the diagnosis by focusing on infection lesions in the lung. The dual-sampling strategy is implemented to address the imbalanced size distribution between COVID-19 and CAP. To segment the lungs, the authors utilized a VB-Net toolkit [76], and for possible lesions regions, auto contouring was performed.

The backbone of the architecture is based on a 3D extended version of the Residual Network, specifically ResNet34 [77]. The utilization of the attention mechanism is widely observed and can be classified into two main categories:activation-based attention [78], [79], and gradient-based attention [80], [81]. The activation-based attention module is commonly incorporated into the network architecture to refine hidden feature maps during training, which enables the network to concentrate on significant areas.

The channel-wise attention assigns weights to each channel in the feature maps [79], while the position-wise attention produces importance heatmaps for each pixel of the feature maps [78], [82].

On the other hand, gradient-based attention methods, such as CAM [80] and Grad-CAM [81], identify important regions that affect network predictions. These methods are typically used offline to provide model interpretability during inference.

Recent studies [83], [84] have highlighted the potential of gradient-based methods for improved localization and their applicability as online modules during training. Building upon this, the authors of this research have extended the gradient-based attention module to create an online trainable component specifically designed for 3D inputs. This proposed attention module incorporates segmented pneumonia infection regions, enabling the network to reach conclusions based on these specific areas of infection. By integrating this attention mechanism, the network can effectively focus on relevant regions and enhance its performance in localizing and classifying pneumonia infections within 3D data.

Zhou *et al.* [85] proposed UNet++, an enhanced version of U-Net with a nested convolutional structure between the encoding and decoding paths. This design improves segmentation performance but also increases the complexity and training requirements.

TABLE VIII: Public Datasets

| dataset | Type | Structure | Ground truth | Link |
|---|---|---|---|---|
| COVID-19 CT Lung and Infection Segmentation Dataset [89] | CT | 20 covid-19 patients | Lung lobs / infection | zenodo.org/record/3757476 |
| CovID-19 CT segmentation dataset [90] | CT | 100 axial CT images | Ground-glass / consolidation / pleural effusion | http://medicalsegmentation.com/covid19/ |
| CovID-19 Radiography Database [91], [92] | X-ray | 3616 Covid-19 cases<br>10,192 Normal cases<br>6012 Lung Opacity<br>1345 Viral Pneumonia image<br>Corresponding lung mask | Covid-19 / Normal / Lung Opacity (Non-CoviD lung infection) / Viral Pneumonia / lung masks | www.kaggle.com/tawsifurrahman/covid19-radiography-database |
| covidx Dataset [25] | X-ray | 16,490 COVID-19 images over 2,800 patients | Normal / COVID-19 / Pneumonia | https://github.com/lindawangg/CoVID-Net/blob/master/docs/COVIDx.me |
| COVID-19 Chest X-Ray Dataset [93] | X-ray | 6504 images /masks | COVID-19 / fungal pneumonia / viral pneumonia / Pneumocystis pneumonia / bacterial pneumonia / Chlamydophila pneumonia | darwin.v7labs.com/v7-labs/covid-19-chest-x-ray-dataset |
| CoronaCases [94] | CT | 10 confirmed COVID-19 cases | - | coronacases.org/ |
| COVID-CT-Dataset [95] | CT | 349 CT images<br>216 patients | - | https://github.com/UCSD-AI4H/COVID-CT |
| MosMedData [96] | CT | Over 1,500 CT scans | - | mosmed.ai |

Chen *et sl.* [86] employed a UNet++ based segmentation model to obtain segmented lesions, which were then used to predict the final label of COVID-19 or non-COVID-19.

ResNet-50 was used as the backbone of UNet++, and all pre-training parameters of ResNet-50 were loaded into UNet++. ResNet-50 was pretrained on the ImageNet dataset. The AI results reduced radiologists' reading time by 65%.

Amyar et al. [87] introduced the system that comprises three main components: a single encoder for feature extraction, decoders for image segmentation, and a multi-layer perceptron for classification.

Gao *et al.* [88] constructed a dual-branch combination network (DCN) that can simultaneously perform individual-level classification and lesion segmentation. The proposed method integrates a novel lesion attention module to combine intermediate segmentation results and enhance the focus of the classification branch on the lesion areas.

The classification branch of the proposed method is based on ResNet-50, which includes four residual blocks. The accuracy achieved by the proposed method for slice-level classification and individual-level classification was 95.99% and 96.74%, respectively. Moreover, the dice score obtained for lung and lesion segmentation were 99.11% and 83.51%, respectively.

## V. PUBLIC IMAGING DATASETS FOR COVID-19

The acquisition of data plays a fundamental task in the initial stages of developing machine learning methodologies for COVID-19 applications. It is essential to assemble a diverse and representative dataset that encompasses a wide range of cases to facilitate the training, validation, and testing of models. Although there exist publicly accessible datasets comprising CT or X-ray images for lung diseases, they may not consistently encompass an adequate number of COVID-19 cases. Consequently, researchers have actively undertaken the task of amassing COVID-19-specific datasets to facilitate the creation and assessment of machine learning models dedicated to the detection, segmentation, and diagnosis of COVID-19. These datasets can be classified into two main categories: (1) datasets that contain both labels and masks, encompassing various pulmonary manifestations such as lung, GGO, consolidation, and other types of lesions. These datasets are intended for tasks involving segmentation, as well as a combination of segmentation and classification, and (2) datasets that comprise images with associated labels, specifically pertaining to COVID-19, viral pneumonia, and other types of pneumonia. These datasets are specially curated for classification tasks. Examples of datasets in the first category include COVID-19 CT Lung and Infection Segmentation Dataset [89], COVID-19 CT Segmentation Dataset [90], COVID-19 Radiography Database [91], [92], and COVID-19 Chest X-Ray Dataset [93] while the second category includes COVIDx Dataset [25], CoronaCases [94], COVID-CT-Dataset [95], and MosMedData [96].Table VIII enumerates the most commonly used public datasets.

## VI. Discussion and Future Work

Despite the progress made in various areas related to COVID-19, there is still much research that needs to be conducted in the future to address the challenges and limitations of current approaches. The following paragraphs detail several potential avenues for future research that could enhance the accuracy, generalizability, and effectiveness of methods for COVID-19 diagnosis, segmentation, follow-up, and treatment. COVID-19 classification is a arduous task that requires the development of accurate and robust models.The scarcity of extensively annotated datasets on a large scale presents a significant obstacle. Most of the existing datasets are small in size and have limited annotations, which makes it difficult to train accurate models. Another challenge is the lack of generalizability of models due to variations in acquisition protocols and imaging devices. Future research can focus on developing large-scale annotated datasets, transfer learning [97] techniques, multimodal(CXR and CT scans) data integration, explainable AI techniques [98], [99], and severity classification to improve the accuracy and generalizability of models. Furthermore, adding other information not only the images, like medical and pathological History, help the model to perform more accurately.

Future research can focus on developing other architecture to enhance the accuracy and generalizability of models.

Follow-up of COVID-19 patients can be performed through a variety of methods such as clinical evaluations, laboratory testing, and imaging studies. Several studies have reported the long-term sequelae of COVID-19, including pulmonary fibrosis, cardiac dysfunction, and neurological complications. Thus, follow-up is essential to identify and manage these complications. Several future research directions can be pursued to improve the monitoring of COVID-19 patients. One of the main directions is the development of standardized protocols for follow-up. Standardized protocols can ensure that patients receive consistent and appropriate care, regardless of the healthcare setting. Moreover, developing a centralized data repository for COVID-19 patients can facilitate the exchange of patient information between healthcare providers, thereby improving continuity of care.

## VII. Conclusion

The advent of the COVID-19 pandemic has engendered an urgent and critical demand for the development of efficacious and accurate methodologies aimed at segmenting and classifying COVID-19-infected regions from medical images. In this scholarly article, we have undertaken a comprehensive analysis of intelligent imaging platforms and cutting-edge research endeavors employing two prominent modalities of medical imaging, namely computed tomography (CT) and X-ray imaging. Our primary objective has been to meticulously examine the latest advancements and groundbreaking contributions in this domain, with a particular emphasis on the techniques and algorithms proposed for achieving precise segmentation and classification of COVID-19-infected regions in medical images. Through our rigorous analysis, we have identified several promising approaches that hold considerable potential for augmenting the efficiency and dependability of COVID-19 detection and diagnosis, which assumes paramount importance within the current healthcare milieu.

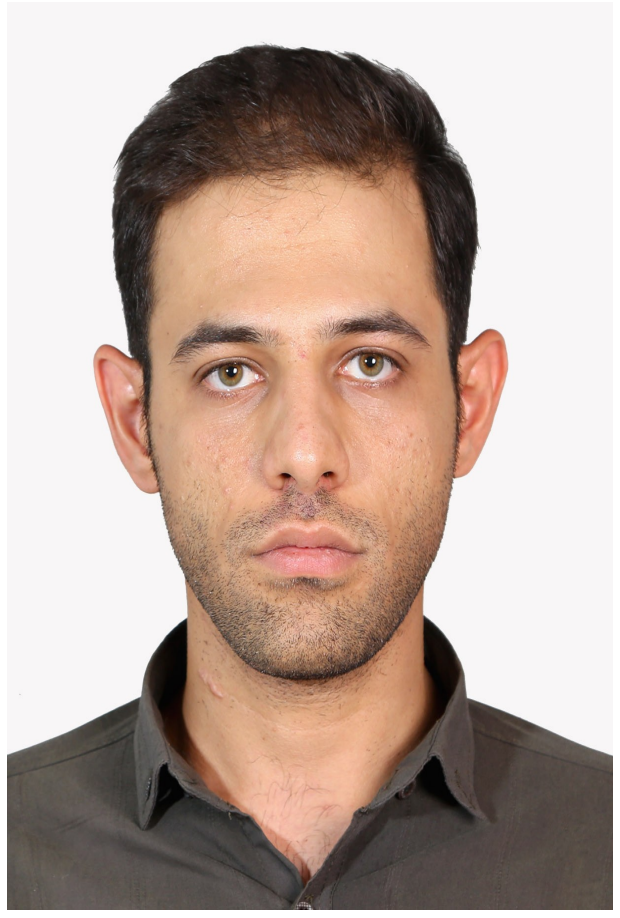

Sayed Amir Mousavi Mobarakeh holds a B.Sc. in Electrical Engineering from Isfahan University (2019) and an M.Sc. from Shiraz University of Technology (2023). Currently pursuing a Ph.D. in Neuroscience at UPJV, France, his research integrates medical imaging, machine learning, and innovative architectural design for diverse medical applications. During his master's, he focused on COVID-19 CT image segmentation, showcasing dedication to addressing critical healthcare challenges. Mousavi Mobarakeh's academic trajectory highlights his commitment to advancing neuroscience and medical engineering, positioning him as a promising researcher at the intersection of technology and healthcare.

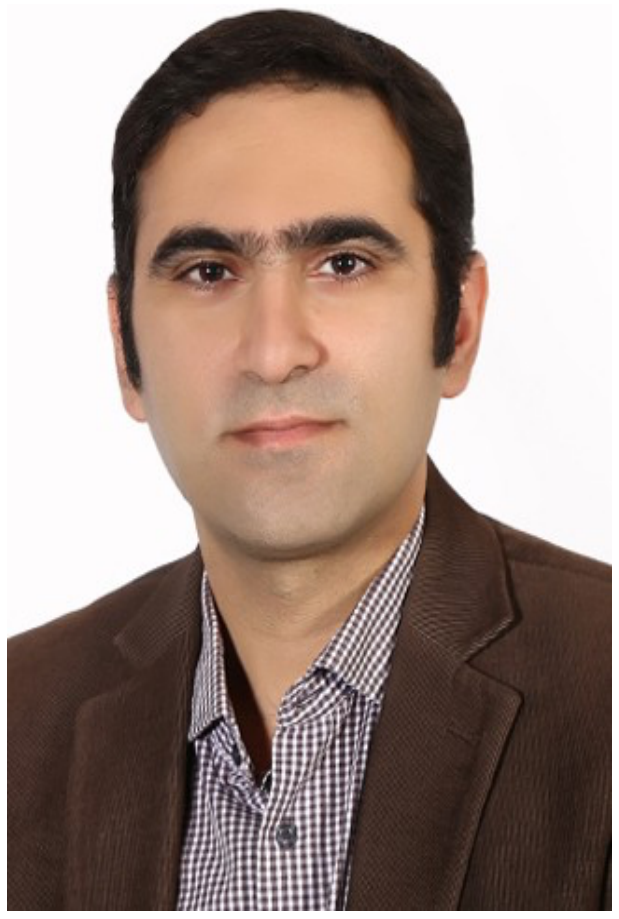

Kamran Kazemi received his B.Sc. and M.Sc. degrees in Electrical engineering from Ferdowsi University of Mashhad and K. N. Toosi University of Technology, Tehran in 2000 and 2002, respectively. He performed his Ph.D. degree in Electrical Engineering and Biomedical Engineering as cooperation between K. N. Toosi University of Technology, Tehran, Iran and University of Picardie Jules Verne (UPJV) Amiens, France. Currently he is Associate Professor in department of Electrical and Electronics Engineering, Shiraz University of Technology, Shiraz, Iran. His present interests are pattern recognition, image processing and machine learning.

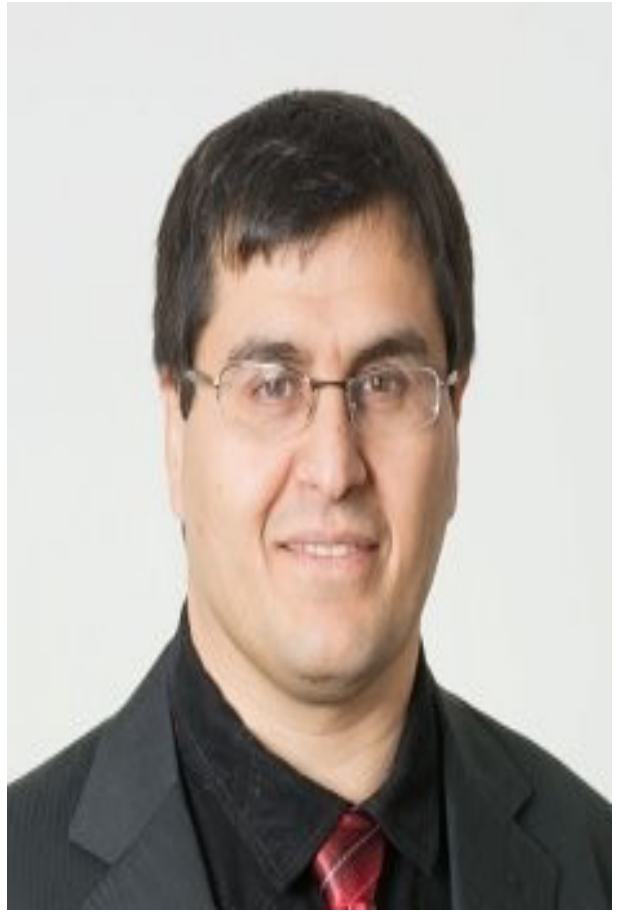

Ardalan Aarabi received his B.Sc. degree in Electrical Engineering from Shiraz University in 1995, followed by an M.Sc. degree in Biomedical Engineering from Iran University of Science and Technology in 1999. He completed his Ph.D. in Neuroscience from the University of Picardy Jules Verne (UPJV, France) in 2007. From 2008 to 2012, he served as a research associate at the University of Manitoba, the University of Minnesota, and the University of Pittsburgh. Currently, he holds the position of Associate Professor at the Faculty of Medicine, University of Picardy Jules Verne. His research interests encompass medical signal and image processing, computational modeling of brain dynamics, and the analysis of functional and structural connectivity in brain networks.

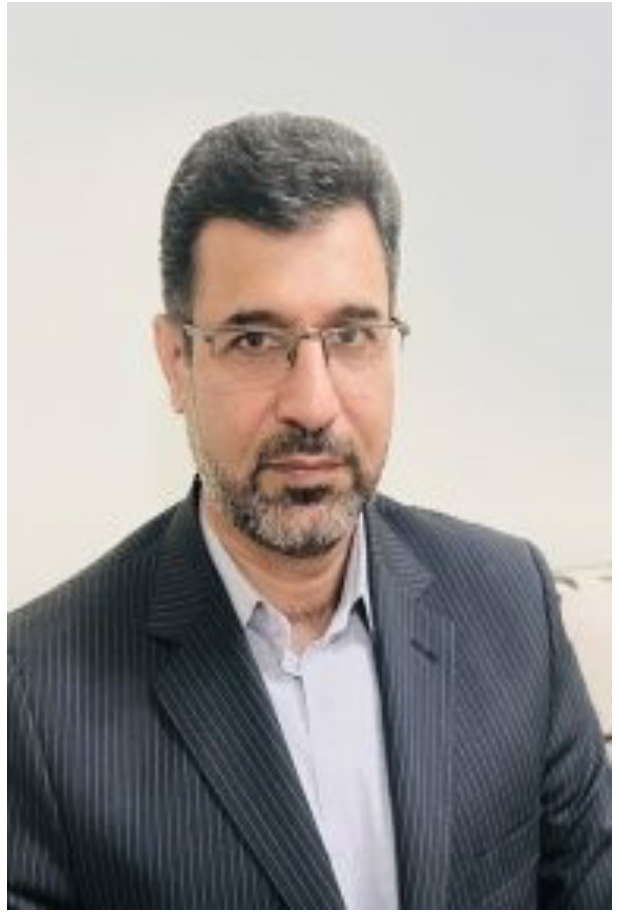

Habibollah Danyali received the B.Sc. degree in electrical engineering from the Isfahan University of Technology, Isfahan, Iran, in 1991, the M.Sc. degree in electrical engineering from Tarbiat Modares University, Tehran, Iran, in 1993, and the Ph.D. degree in computer engineering from the University of Wollongong, Wollongong, NSW, Australia, in 2004. He is currently a Professor with the Department of Electrical Engineering, Shiraz University of Technology, Shiraz, Iran. His research interests include scalable image and video coding, remote sensing, and machine learning.